\documentclass[10pt,conference]{IEEEtran}
\IEEEoverridecommandlockouts
\usepackage[utf8]{inputenc}
\usepackage[T1]{fontenc}
\usepackage{cite}
\usepackage{amsmath,amssymb,amsfonts}
\usepackage{graphicx}
\usepackage{textcomp}
\usepackage{xcolor}
\usepackage{booktabs}
\usepackage{array}
\usepackage{url}
\usepackage{multirow}
\usepackage{enumitem}
\usepackage{caption}
\usepackage{subcaption}
\usepackage{tikz}
\usetikzlibrary{arrows.meta,positioning,shapes.geometric,fit,calc,backgrounds}
\usepackage[hidelinks]{hyperref}

\usepackage{flushend}

\begin{document}

\title{Preparing an AI-Augmented SIEM for the EU Cyber Resilience Act: A Practitioner Case Study}

\author{%
\IEEEauthorblockN{Georgios Koutidis\IEEEauthorrefmark{1},
Nikolaos Kekatos\IEEEauthorrefmark{1},
Marina Korgiala Karyda\IEEEauthorrefmark{2},
Alexios Lekidis\IEEEauthorrefmark{2}, and
Tom Nianios\IEEEauthorrefmark{1}}
\IEEEauthorblockA{%
\IEEEauthorrefmark{1}\textit{Clone Systems}, Larnaca, Cyprus\\
\{gkoutidis, nkekatos, tnianios\}@clone-systems.com\\[2pt]
\IEEEauthorrefmark{2}\textit{University of Thessaly}, Larissa, Greece\\
korgialamarina@gmail.com, alekidis@uth.gr}%
}
\maketitle
{\renewcommand{\thefootnote}{}\footnotetext{\copyright~2026 IEEE. Personal use of this material is permitted. Permission from IEEE must be obtained for all other uses, in any current or future media, including reprinting/republishing this material for advertising or promotional purposes, creating new collective works, for resale or redistribution to servers or lists, or reuse of any copyrighted component of this work in other works. This work has been accepted for publication at the 2026 IEEE International Conference on Cyber Security and Resilience (IEEE CSR 2026).}}

\begin{abstract}
The EU Cyber Resilience Act (CRA), Regulation (EU)~2024/2847, makes product cybersecurity a lifecycle obligation for products with digital elements on the EU market: risk assessment, vulnerability handling, conformity documentation, and Article~14 incident- and vulnerability-reporting readiness must be operational before market placement. Small and medium-sized enterprises that build security products are doubly exposed, since their products are in scope while their customers expect them to be exemplary. This case study documents a CRA preparedness pilot for one such product, \emph{SEUXDR}, an AI-augmented security monitoring product with a large-language-model active-response component, on the open-source CYBERFORT platform. We contribute a reproducible six-step recipe (Scope and Classify, Asset Registration, Produce Evidence, Map to CRA, Gap and Actions, Audit Pack), two end-to-end traceability threads, and a pilot snapshot tracing product risks through baseline and AI-specific controls and policies to CRA objectives. It offers practitioners a replicable starting point for translating CRA legal text into operational preparedness for incident response, vulnerability reporting, and conformity assessment.
\end{abstract}

\begin{IEEEkeywords}
Cyber Resilience Act, preparedness, situational awareness, incident response readiness, conformity preparation, SME cybersecurity, AI-augmented SIEM, case study.
\end{IEEEkeywords}

\section{Introduction}\label{sec:intro}
The EU Cyber Resilience Act (CRA) establishes mandatory horizontal cybersecurity requirements for products with digital elements made available on the EU market~\cite{eu_cra_2024}. It entered into force in December~2024; the Article~14 vulnerability- and incident-reporting obligations apply from September~2026, and the essential cybersecurity requirements (Annex~I) from December~2027. Manufacturers must document a cybersecurity risk assessment, embed it in the product lifecycle, operate a vulnerability-handling process, track actively exploited vulnerabilities and severe incidents, and prepare conformity documentation including the EU declaration of conformity. CRA compliance is therefore not a one-off checklist but an ongoing \emph{preparedness} problem: continuous visibility over assets, risks, controls, evidence, and incident-response artefacts must be in place before the regulation applies.

Small and medium-sized enterprises (SMEs) that build security products are doubly exposed. Their products are in scope, since a Security Information and Event Management (SIEM) system, an Extended Detection and Response (XDR) tool, or an Endpoint Detection and Response tool is itself software with digital elements, and many fall under Annex~III as important products, Class~I in the case of SIEM systems. At the same time, customers expect security vendors to lead on security, yet those same SMEs typically run on small governance, risk, and compliance (GRC) teams with spreadsheets and ad-hoc evidence collection; recent work by the EU Agency for Cybersecurity (ENISA) documents how low CRA awareness still is among small manufacturers~\cite{szedlak2024cra}. A key observation motivates this work: an open-source SIEM such as Wazuh\footnote{Wazuh, the open source security platform: \url{https://wazuh.com/}.} detects, correlates, and responds to threats, but produces none of the conformity artefacts the CRA requires. The distance between a working SIEM and a CRA-ready product is precisely the documentation, traceability, and reporting workflow this case study builds, and bridging it is tractable only once someone has done it end-to-end on a real product.

This case study documents that exercise for \emph{SEUXDR} (NextGen SIEM with AI Remediation), an AI-augmented SIEM/XDR product developed by Clone Systems, on the open-source CYBERFORT platform~\cite{cyberfort_platform_paper},\footnote{Platform: \url{https://cyber-fort.eu/}; open-source code: \url{https://github.com/CloneSystems/cyberfort_v3}; hosted instance, free after registration: \url{https://access.cyber-fort.eu/}.} built specifically for SMEs preparing for the CRA. The central idea is the \emph{compliance chain}, a traceable path linking each CRA obligation through framework controls, product risk, policies, technical evidence, and assessment status all the way to the technical-documentation file and EU declaration of conformity, introduced in~\cite{cyberfort_platform_paper} and applied end-to-end on a real product here. The paper is intentionally narrative and screenshot-led: it documents what we entered, what the platform produced, what we changed, and what we wish we had known. At a high level, the contributions are: (i)~a reproducible six-step CRA preparedness recipe aligned with SME engineering reality; (ii)~a concrete worked example on an AI-augmented security product (SEUXDR), spanning both conventional SIEM exposures and Large Language Model (LLM)-driven active-response risks; (iii)~an end-to-end traceability matrix from product risks through controls and policies to CRA objectives, including Article~14 reporting-readiness artefacts; (iv)~a pilot snapshot with quantitative coverage and the CRA-oriented audit-pack reports; and (v)~case-study lessons, advantages, and limitations for AI-augmented security products under the CRA. The remainder covers CRA background and positioning (\S\ref{sec:related}), pilot setup (\S\ref{sec:pilot-setup}), the recipe (\S\ref{sec:methodology}), results (\S\ref{sec:snapshot}), discussion (\S\ref{sec:discussion}), and conclusion (\S\ref{sec:conclusion}).

\section{Background and Related Work}\label{sec:related}
Beyond the obligations and timelines of \S\ref{sec:intro}, the CRA's Annex~I states the essential requirements, Annex~III lists important products, and conformity routes range from manufacturer-internal control to third-party assessment by product class. Several resources translate this text into engineering vocabulary: ENISA's CRA-to-standards mapping~\cite{enisa_cra_mapping} relates Annex~I to harmonised standards, NIST~CSF~2.0~\cite{nist_csf2} and IEC~62443~\cite{iec62443} supply reusable control catalogues, and the companion paper~\cite{cyberfort_platform_paper} describes the CYBERFORT platform and the compliance-chain model in full.

Two gaps remain. First, the SME-readiness literature is largely diagnostic: ENISA-aligned surveys report low awareness among small manufacturers~\cite{szedlak2024cra} but not what a manufacturer must actually produce before review. Second, standards crosswalks describe \emph{what} maps to \emph{what}, not \emph{how} a small team assembles, links, and evidences those mappings on a concrete product, still less one embedding AI in its decision loop. End-to-end practitioner case studies on a real product with AI components remain rare, the space this paper occupies.

The novelty is best stated against a traditional SIEM. A conventional SIEM/XDR such as Wazuh is a security \emph{control}: it ingests logs, correlates events, scores severity, and can drive active response, but it is not a compliance instrument. It does not register the product as a CRA asset, map its residual risks to Annex~I obligations, maintain approved policies and a provenance-tracked evidence index, or emit the technical documentation and EU declaration of conformity. The approach here wraps such a SIEM in a GRC layer that produces exactly those artefacts and the end-to-end traceability (the compliance chain) an auditor consults, so an existing detection tool becomes a CRA-ready product with one navigable source of truth from each Common Vulnerabilities and Exposures (CVE) entry and product risk to the CRA objective it touches, without bespoke tooling or a dedicated compliance team.

\section{Pilot Setup: SEUXDR and CRA Scope}\label{sec:pilot-setup}
This section fixes the two things that drive every later artefact: what the product is (\S\ref{ssec:arch}--\S\ref{ssec:nontrivial}) and what the CRA says about it (\S\ref{ssec:cra-class}).

Clone Systems CY LTD is a Managed Security Service Provider based in Larnaca, Cyprus; it ships SEUXDR and acts as the CRA manufacturer. The pilot began in October~2025 with six engineers (product, security, and GRC) executing the six recipe steps described in Section~\ref{sec:methodology}.

\subsection{Architecture and Operator Surface}\label{ssec:arch}
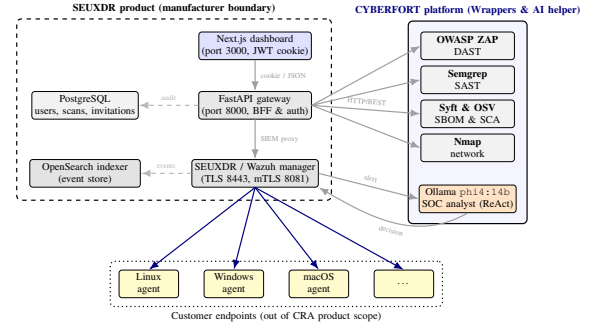
\begin{figure}[tb]
\centering
\resizebox{0.85\columnwidth}{!}{%
\begin{tikzpicture}[
  font=\scriptsize,
  node distance=2.5mm,
  box/.style       ={rectangle,rounded corners=2pt,draw,align=center,
                     minimum height=6.5mm,inner sep=2pt},
  ui/.style        ={box,fill=blue!12,minimum width=24mm},
  gw/.style        ={box,fill=gray!18,minimum width=24mm},
  siem/.style      ={box,fill=gray!22,minimum width=26mm},
  ai/.style        ={box,fill=orange!22,minimum width=22mm},
  db/.style        ={box,fill=gray!10,minimum width=24mm},
  agent/.style     ={box,fill=yellow!25,minimum width=14mm},
  scanner/.style   ={box,fill=gray!10,minimum width=22mm},
  arrow/.style     ={-Latex,thick,gray!75},
  mtls/.style      ={-Latex,thick,blue!50!black},
  internal/.style  ={-Latex,thick,gray!55,dashed},
]
\node[ui]    (front)  at (-0.5, 4.2) {Next.js dashboard\\\scriptsize(port 3000, JWT cookie)};
\node[gw]    (gw)     at (-0.5, 2.8) {FastAPI gateway\\\scriptsize(port 8000, BFF \& auth)};
\node[db]    (pg)     at (-4.5, 2.8) {PostgreSQL\\\scriptsize users, scans, invitations};
\node[siem]  (siemmgr) at (-0.5, 1.2) {SEUXDR / Wazuh manager\\\scriptsize(TLS 8443, mTLS 8081)};
\node[siem]  (idx)     at (-4.5, 1.2) {OpenSearch indexer\\\scriptsize(event store)};

\node[scanner] (sc1) at (4.5, 4.2) {\textbf{OWASP ZAP}\\\scriptsize DAST};
\node[scanner] (sc2) at (4.5, 3.4) {\textbf{Semgrep}\\\scriptsize SAST};
\node[scanner] (sc3) at (4.5, 2.6) {\textbf{Syft \& OSV}\\\scriptsize SBOM \& SCA};
\node[scanner] (sc4) at (4.5, 1.8) {\textbf{Nmap}\\\scriptsize network};
\node[ai]      (ai)  at (4.5, 0.6) {Ollama \texttt{phi4:14b}\\\scriptsize SOC analyst (ReAct)};

\begin{scope}[on background layer]
  \node[draw,thick,rounded corners=3pt,fill=blue!4,fit=(sc1)(sc2)(sc3)(sc4)(ai),
        inner sep=2.5mm,
        label={[font=\scriptsize\bfseries,blue!50!black,anchor=south]
        above:CYBERFORT platform (Wrappers \& AI helper)}] (cfbox) {};
\end{scope}

\node[agent] (a1) at (-3.0, -1.4) {Linux\\agent};
\node[agent] (a2) at (-1.0, -1.4) {Windows\\agent};
\node[agent] (a3) at ( 1.0, -1.4) {macOS\\agent};
\node[agent] (a4) at ( 3.0, -1.4) {\dots};

\draw[arrow]    (front.south) -- node[right,font=\tiny]{cookie / JSON} (gw.north);
\draw[internal] (gw.west)    -- node[above,font=\tiny]{audit}        (pg.east);
\draw[arrow]    (gw.east)    -- node[above,font=\tiny,sloped]{HTTP/REST} (sc3.west);
\draw[arrow]    (gw.east)    -- (sc1.west);
\draw[arrow]    (gw.east)    -- (sc2.west);
\draw[arrow]    (gw.east)    -- (sc4.west);
\draw[arrow]    (gw.south)   -- node[right,font=\tiny]{SIEM proxy}    (siemmgr.north);
\draw[internal] (siemmgr.west) -- node[above,font=\tiny]{events}      (idx.east);
\draw[arrow]    (siemmgr.east) -- node[above,font=\tiny,sloped]{alert} (ai.west);
\draw[arrow]    (ai.south) to[out=-150,in=-30] node[below,font=\tiny,sloped]{decision} (siemmgr.south east);
\foreach \a in {a1,a2,a3,a4}{
  \draw[mtls] (siemmgr.south) -- (\a.north);
}

\begin{scope}[on background layer]
  \node[draw,thick,rounded corners=3pt,dashed,fit=(front)(gw)(pg)(siemmgr)(idx),
        inner sep=3mm,
        label={[font=\scriptsize\bfseries,anchor=south]
        above:SEUXDR product (manufacturer boundary)}] {};
  \node[draw,thick,rounded corners=3pt,dotted,fit=(a1)(a4),
        inner sep=2mm,
        label={[font=\scriptsize,anchor=north]
        below:Customer endpoints (out of CRA product scope)}] {};
\end{scope}
\end{tikzpicture}%
}
\caption{SEUXDR system view: SEUXDR manufacturer boundary on the left, CYBERFORT tooling block (Wrappers \& AI helper) on the right, customer-side agents below. DAST: dynamic application security testing; SAST: static application security testing; SBOM: software bill of materials; SCA: software-composition analysis.}
\label{fig:seuxdr-arch}
\end{figure}

SEUXDR is delivered as a multi-tenant web application whose SIEM workspace is organised into six tabs (\emph{Overview}, \emph{Alerts}, \emph{Agents}, \emph{Remediation}, \emph{Installer}, and \emph{MITRE ATT\&CK}) covering host-based intrusion detection, log analysis, and active response across Linux, Windows, and macOS endpoints. Figure~\ref{fig:seuxdr-arch} shows the system view derived from the product's two open-source repositories, with two trust boundaries: the SEUXDR manufacturer scope on the left, and the CYBERFORT platform (containerised scanners plus a local Ollama \texttt{phi4:14b} security operations centre (SOC) analyst) on the right. The endpoint agents (dotted frame) communicate with the manager over mutual Transport Layer Security (mTLS) using a ``smart manager / dumb agent'' active-response pattern: the manager generates OS-specific commands that agents run through a whitelisted executor.

\paragraph{Endpoints / Agents} Self-updating agents on Windows, Linux, and macOS provide log shipping and file-integrity monitoring, registering via mTLS-anchored identities. The \emph{Installer} tab generates a one-shot installer per platform; registered endpoints appear under the \emph{Agents} tab, each with a \emph{Deactivate} action that revokes its certificate.

\paragraph{Manager} The manager runs Wazuh with an OpenSearch backend for alert correlation, storage, and severity scoring. The \emph{Alerts} dashboard (not pictured) exposes filtering and severity views, and the \emph{MITRE ATT\&CK} tab lets analysts pivot from a triggered alert into the corresponding adversary technique.

\paragraph{AI Actions} An LLM-driven agent (a locally hosted model, accessed via Ollama\footnote{Ollama, run large language models locally: \url{https://ollama.com/}.}) consumes Wazuh alerts and selects active-response actions using a ReAct-style reason-then-act loop~\cite{yao2023react}, falling back to rule logic when the model is unavailable or the action lies outside its sanctioned playbook scope. A \emph{Remediation} tab (not pictured) lets the operator review AI behaviour, trigger manual remediation, and audit per-agent histories, with drill-downs exposing the Active Response feed and Failed Alerts stream for human-in-the-loop oversight. TLS secures the web and application programming interface (API) surface and mTLS the agent channel; authentication uses JSON Web Tokens (JWT) with role-based access control (RBAC), audit-logged throughout.

\subsection{Threat Model and Why the Product Is Non-Trivial}\label{ssec:nontrivial}
SEUXDR is itself a security product placed on the EU market by Clone Systems acting as manufacturer, and its AI Actions component raises CRA-relevant risks (unsafe actions, model drift, privileged automation) that go beyond the baseline SIEM threat model and require additional controls from the Secure Controls Framework (SCF)~\cite{scf_aat}, specifically its \emph{AI \& Autonomous Technologies} (AAT) family. Table~\ref{tab:threat-model} maps the four product components in Fig.~\ref{fig:seuxdr-arch} to the CRA-relevant threats they expose and the Annex~I obligation each triggers.

The threats in Table~\ref{tab:threat-model} were enumerated method-by-method rather than ad hoc: we decomposed SEUXDR into its four trust-bounded components (Fig.~\ref{fig:seuxdr-arch}) and applied three established methods per component: STRIDE~\cite{shostack2014} for the dashboard, gateway, and manager data flows; the OWASP Top~10~\cite{owasp_top10} for the web and API surface; and MITRE ATT\&CK~\cite{mitre_attack} techniques already encoded in the Wazuh ruleset for the agent and active-response channel. Threats specific to the AI Actions component (emergent or unsafe autonomous actions, model drift, privileged-action escalation, prompt manipulation), which none of the three baseline methods cover, were drawn from the SCF AAT family. Each enumerated threat was then mapped to the Annex~I obligation it triggers, and that mapping seeds the Step~4 risk register.

\begin{table}[tb]
\caption{SEUXDR component-to-threat-to-Annex~I mapping}
\label{tab:threat-model}
\centering
\scriptsize
\setlength{\tabcolsep}{3pt}
\renewcommand{\arraystretch}{0.95}
\begin{tabular}{@{}>{\raggedright\arraybackslash}p{0.18\columnwidth}>{\raggedright\arraybackslash}p{0.37\columnwidth}>{\raggedright\arraybackslash}p{0.37\columnwidth}@{}}
\toprule
\textbf{Component} & \textbf{CRA-relevant threats} & \textbf{Annex~I obligation} \\
\midrule
Dashboard / gateway (Next.js \& FastAPI) & Broken authentication, session fixation, API abuse, insecure direct object reference & Annex~I~(2)(d) protection from unauthorised access; (2)(e) confidentiality \\
SIEM manager (Wazuh \& indexer) & Tampered logs, lost integrity of evidence, supply-chain compromise of ruleset & Annex~I~(2)(f) integrity of data; Part~II (vulnerability handling) \\
AI SOC analyst (Ollama \texttt{phi4:14b}) & Emergent / unsafe autonomous actions, model drift, privileged-action escalation, prompt manipulation & Annex~I~(2)(i) minimise negative impact; Art.~13 (manufacturer duties) \\
Endpoint agents (mTLS) & Credential theft, lateral movement via active-response channel, replay of remediation commands & Annex~I~(2)(j) limit attack surfaces; (2)(k) reduce impact of incidents \\
\bottomrule
\end{tabular}
\end{table}

\subsection{CRA Classification and Resource Footprint}\label{ssec:cra-class}
SEUXDR is registered in CYBERFORT as software with digital elements. Under CRA Annex~III~\cite{eu_cra_2024}, SIEM systems fall under \emph{Important Products with Digital Elements, Class~I}. Clone Systems acts as the manufacturer (Art.~13). The conformity route is internal control supplemented by the EU declaration of conformity once the technical documentation is complete (Art.~32). Internal control under Art.~32 is the manufacturer-internal route applicable here; products falling under Annex~III Class~II or the critical-products list would instead require third-party assessment. Within CYBERFORT this is selected during scope assessment and locked in for downstream gap analysis and reporting.

\paragraph{Pilot goals} The pilot does not aim to declare conformity. It targets four CRA obligation families (secure-by-design and secure-by-default, vulnerability handling, conformity evidence, and Article~14 reporting readiness) and produces a product-boundary dossier, a control-mapping matrix, a policy uplift plan, a risk-treatment mapping, and a management decision pack.

\paragraph{Resource footprint} Because the CRA touches the whole lifecycle, an SME must also know what the preparedness layer costs to run. Compute is dominated by the local Ollama \texttt{phi4:14b} SOC analyst, which runs on a single workstation-class GPU host and is invoked once per correlated alert rather than once per raw event (with rule-based fallback), so detection latency is bounded by Wazuh correlation, not model inference; the five containerised scanners run on demand in Step~3, not continuously. Storage is dominated by the OpenSearch event store sized to the retention policy, while the evidence index (software bill of materials, or SBOM, plus configuration exports and audit-event samples) and the compliance-chain graph (1{,}775 edges) are small structured documents. Communication is modest: heartbeat and event shipping over mTLS, short HTTP/REST scanner bursts, and a one-shot Article~14 submission. The preparedness layer thus adds governance metadata and on-demand scanning to an existing SIEM rather than a new always-on workload.

\section{Methodology: A Six-Step CRA Preparedness Recipe on CYBERFORT}\label{sec:methodology}
This section operationalises CRA legal obligations into a six-step engineering recipe executed on CYBERFORT: Scope and Classify $\to$ Asset Registration $\to$ Produce Evidence $\to$ Map to CRA $\to$ Gap and Actions $\to$ Audit Pack. Each step produces an artefact that closes one link in the compliance chain so that, by Step~6, every CRA obligation traces back through a policy, a control, a risk, and a product asset; the resulting chain is rendered by the platform's \emph{Compliance Chain~/~Map} view (Fig.~\ref{fig:chain}, shown later in Step~4 once the necessary entities exist). Each step is reproducible from a fresh CYBERFORT install once the framework seed data is loaded.

The remainder of this section follows CYBERFORT's \emph{CRA Start Example} entry-point on SEUXDR, showing for each step what the operator does and what the platform produces.

\subsection{Step 1: Scope and Classify}
The first decision in the recipe is what counts as ``the product''. The scope-assessment workflow elicits the economic-operator role (\emph{Manufacturer}), CRA applicability (\emph{in scope}), Annex classification (\emph{Annex~III, Class~I, SIEM systems}), and conformity route (\emph{internal control supplemented by EU declaration} under Art.~32). The platform seeds the CRA framework with its 42 objectives, control library, and policy templates, locks the classification in, and emits a scope report that becomes the entry point for the technical-documentation file.

\emph{Watch-out.} The most consequential call is the product boundary: we scoped SEUXDR as the manager, AI Actions component, and agents, but \emph{not} customer log sources, host OS, or deployment infrastructure. Because every downstream artefact inherits it, we signed it off with engineering, product, and legal before leaving Step~1.

\subsection{Step 2: Asset Registration}
With the scope locked in, the product, its components, and its sub-assets are created as records in CYBERFORT. SEUXDR is registered as a top-level Software asset, with identity, license, and SBOM fields captured up-front as a stable anchor for downstream evidence. Sub-assets are then attached: the Wazuh manager image, the OpenSearch backend, the agent binaries, the AI Actions container, the playbook repository, and the third-party libraries pulled in via the SBOM. Each sub-asset inherits support period and conformity route from the scope unless overridden.

\emph{Watch-outs.} Sub-assets should be registered at the granularity of the smallest item that can independently fail a CRA obligation: the AI Actions container, for instance, has its own support period and conformity-relevant evidence (model card, playbook hash) distinct from the manager.

\subsection{Step 3: Produce Evidence}\label{ssec:evidence}
Step~3 produces the \emph{Evidence} objects on which the rest of the recipe depends. The \emph{Wrappers} integration layer runs five containerised scanners: Syft (SBOM generation in CycloneDX), OSV Scanner (software-composition analysis (SCA) via CVE lookup against the Syft SBOM), Semgrep (static application security testing (SAST) and policy-as-code over the Wazuh customisations, AI Actions code, and playbooks), OWASP~ZAP (dynamic application security testing (DAST) against the manager web/API), and Nmap (network-segmentation verification), with an on-demand SBOM Generator helper alongside. In parallel, artefacts that already exist outside CYBERFORT (mTLS certificate-issuance procedure, Wazuh ruleset and version pin, audit-log samples, change-management log, AI playbooks and kill-switch test records, prior penetration-test reports) are bulk-loaded into the same Evidence index. Both sources share the same provenance schema (tool, version, scan date, target hash, signed result, owner), so the downstream mapping module does not distinguish between them.

\emph{Watch-outs.} Evidence must be measured against the CRA support period, not the calendar year: an SBOM from an unmaintained release is stale regardless of its date.

\subsection{Step 4: Map to CRA: Risks, Controls, Policies}
The mapping module connects Annex~I obligations to the CYBERFORT control library across six themes; ENISA's CRA-to-standards mapping~\cite{enisa_cra_mapping}, NIST~CSF~2.0~\cite{nist_csf2}, and IEC~62443~\cite{iec62443} are useful starting points for the control library. The six themes, with their main SEUXDR components and the gap the pilot closed for each, were: \emph{identity and access} (JWT, RBAC, and session/role controls; gap: a product-specific access-control standard operating procedure (SOP) and admin hardening); \emph{cryptography and secure communications} (mTLS, certificate pinning, TLS; gap: an approved encryption policy and release-baseline settings); \emph{vulnerability management and updates} (self-updating agents, Wazuh detections, patch workflows; gap: a formalised Product Security Incident Response Team (PSIRT) workflow, service-level agreement (SLA), release evidence, and SBOM routine); \emph{logging and detection} (audit logs, Wazuh/OpenSearch, real-time alerts; gap: defined CRA reporting-evidence fields and a retention strategy); \emph{supply chain and vendors} (OS and Wazuh packages, the Ollama model, Docker images, third-party libraries; gap: a supplier inventory and an SCA/SBOM cadence); and \emph{secure development and testing} (input validation, rate limiting, rule testing, CI/CD; gap: a Secure Software Development Life Cycle (SSDLC) checklist, security-test sign-off, and release gates). Each row of this mapping is realised by linked records under a fixed identifier scheme that carries no meaning beyond grouping: risks are prefixed \texttt{RSK-}, controls carry a control-family prefix (\texttt{IAM-} identity and access, \texttt{AAT-} AI and autonomous technologies, \texttt{CHM-} change management, \texttt{GOV-} governance), and policies \texttt{POL-}. A \emph{Risk} record (e.g.\ \texttt{RSK-102} \emph{emergent properties from AI / autonomous technologies}, severity Critical, treatment Reduce) captures the exposure. A \emph{Control} record (e.g.\ \texttt{AAT-02.2} \emph{AI \& Autonomous Technologies Internal Controls}, Status \emph{Implemented}, linked to Annex~I~(2)(i)) captures the safeguard. A \emph{Policy} record closes the triangle, e.g.\ the \emph{Risk Management Policy} \texttt{POL-30}, created in \emph{Approved} status with the Chief Information Security Officer (CISO) as owner and an effective date set up-front, linked to the CRA framework so it is auto-discovered from the objective checklist.

\begin{figure}[tb]
\centering
\includegraphics[width=\columnwidth]{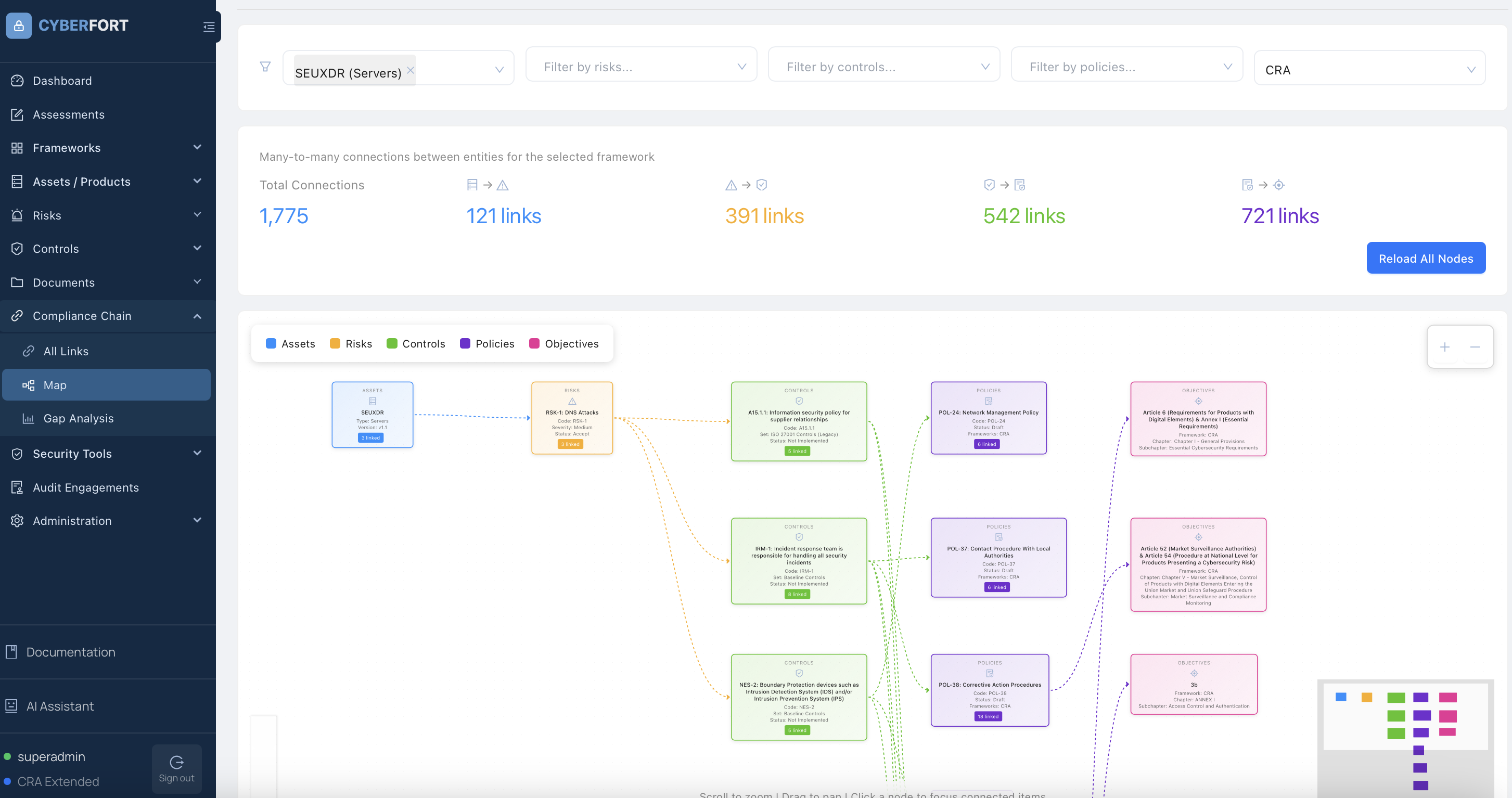}
\caption{The compliance chain as rendered by the CYBERFORT platform for the SEUXDR scope: 1{,}775 many-to-many connections across five colour-coded entity types (Assets to Risks to Controls to Policies to Objectives), with per-link counters on the side and an entity-filter strip across the top. Each node is clickable for evidence drill-down.}
\label{fig:chain}
\end{figure}

Once every Risk--Control--Policy triangle is in place, CYBERFORT's \emph{Compliance Chain~/~Map} view (Fig.~\ref{fig:chain}) renders the resulting graph as a single navigable picture; for SEUXDR it spans 1{,}775 connections across the five entity types, each clickable for evidence drill-down.

\subsection{Step 5: Gap and Actions}
With the chain rendered, Step~5 asks what is missing. The readiness module turns the mapping into a gap list with owner, due date, and closure criterion. A CRA \emph{Conformity} assessment is opened against the SEUXDR asset and the operator walks the questionnaire item by item, attaching the matching policy and evidence record so the compliance chain is complete by close. Each Annex~I objective is then marked \emph{Compliant}, \emph{Partially Compliant}, or \emph{Not Assessed}, and \emph{Generate AI Suggestions} produces recommendations for items not yet fully compliant. Most gaps in the SEUXDR pilot were \emph{documentation} gaps (a control was implemented but the policy was not approved, the evidence was not indexed, or the SLA was not written), the dominant failure mode for security products with strong engineering culture and weak GRC discipline.

\subsection{Step 6: Audit Pack}
Once the corrective-action register has driven each open gap to closure, Step~6 closes the recipe by exporting the seven CRA-oriented reports detailed in Section~\ref{sec:outcomes}. The audit pack is the deliverable that goes to management for the conformity decision and is the input to the internal declaration of conformity.

\section{Results: Threads, Snapshot, and Audit Pack}\label{sec:snapshot}
We report the six-step methodology's output bottom-up: two end-to-end traceability threads make the compliance chain concrete (\S\ref{sec:thread}); the aggregate pilot snapshot then summarises 30 risks, 65 controls, and 27 CRA objectives across the SEUXDR scope (\S\ref{ssec:snapshot}); the audit pack and an Article~14 reporting-readiness rehearsal close the loop (\S\ref{sec:outcomes}).

\subsection{Worked Traceability Threads}\label{sec:thread}
The compliance chain is easier to grasp through one full thread than through any aggregate table. Table~\ref{tab:threads} reproduces two threads side by side, one conventional and one AI-specific. Thread~1 follows the Manager API, the component most exposed to operator interaction: the residual risk of broken or weak authentication (\texttt{RSK-18}) is reduced by a secure-authentication baseline (\texttt{IAM-5}) and mandatory multi-factor authentication (MFA, \texttt{IAM-6}), documented by the Password Policy (\texttt{POL-1}) and evidenced so as to satisfy Annex~I~(2)(d). Thread~2 follows the AI Actions component, whose autonomous remediation (\texttt{RSK-102}) pulls in ten SCF~AAT controls on top of the baseline change-management and privileged-access controls, with operational rather than configurational evidence: AI decision logs, human approval records, a kill-switch test report, and a rollback playbook.

\begin{table}[tb]
\caption{Two worked traceability threads from the SEUXDR matrix: a conventional IAM thread and an AI active-response thread}
\label{tab:threads}
\centering
\scriptsize
\setlength{\tabcolsep}{3pt}
\renewcommand{\arraystretch}{0.95}
\begin{tabular}{@{}p{0.15\columnwidth}p{0.37\columnwidth}p{0.37\columnwidth}@{}}
\toprule
\textbf{Layer} & \textbf{Thread 1: IAM} & \textbf{Thread 2: AI active response} \\
\midrule
Component & Manager API & AI Actions (LLM-driven active response) \\
Risk & \texttt{RSK-18}, broken/weak authentication (High) & \texttt{RSK-102}, emergent properties from AI / autonomous tech.\ (Critical) \\
Controls & \texttt{IAM-5} (secure-auth baseline), \texttt{IAM-6} (mandatory MFA: dashboard, API, admin) & \texttt{AAT-01} (AI governance), \texttt{AAT-02.1/02.2} (AI risk mapping \& internal controls), \texttt{GOV-3} (controls monitoring), \texttt{IAM-4} (privileged-action approval), \texttt{CHM-1/2/3} (playbook change tracking) \\
Policy & \texttt{POL-1} (Password Policy) & \texttt{POL-30} (Risk Mgmt.), \texttt{POL-11} (Change Mgmt.), \texttt{POL-21} (Information Security) \\
CRA objective & Annex~I~(2)(d), unauthorised-access protection & Ch.~II Art.~13 (manufacturer duties, internal processes); Annex~I~(2)(i) \\
Evidence & Auth/RBAC config export; login audit events; password/MFA screenshots & AI decision logs; human approval records; kill-switch test report; rollback playbook \\
\bottomrule
\end{tabular}
\end{table}

\subsection{Pilot Snapshot: Coverage and Severity}\label{ssec:snapshot}
The SEUXDR pilot snapshot, as exported from CYBERFORT, is as follows. SEUXDR is registered as a Software asset of CRA criticality Annex~III, Class~I, with Clone Systems as manufacturer. The pilot identified 30 risks (from 104 templates, spanning 8 of 12 categories), mapped 65 controls (55 from the CYBERFORT Baseline of 132 controls plus 10 from the SCF~v2025.1 AAT family), engaged 25 of 48 policies, and addressed 27 of 42 CRA objectives (including Annex~I and vulnerability handling). Two patterns stand out. The AI and autonomous-operations focus area carries the smallest risk count (4 risks) yet ties application and API security for the largest control set (14 controls), so the cost of safe autonomous operation is paid in control overhead. The severity profile (12 critical, 14 high, 4 medium) skews high, the expected profile for a security product registered as manufacturer, because the product itself is part of the attack surface.

\subsection{Audit Pack and Reporting Readiness}\label{sec:outcomes}
At the end of the pilot, CYBERFORT auto-generates seven CRA-oriented artefacts from the compliance chain: six reports that constitute the audit pack (the CRA assessment, scanning-tool, risks, controls, and policy reports, and the Annex~I objective checklist), plus the EU declaration of conformity, which remains in progress pending legal review of the manufacturer-internal control assertion under Art.~32.

The pilot success criteria, set up-front, were: traceability across the full chain, approved policies, a validated vulnerability-handling workflow, and a reporting-readiness rehearsal under simulated Art.~14 conditions. All four were met for the in-scope SEUXDR boundary.

\subsubsection*{Article~14 dry-run}
As the final readiness check we rehearsed an Art.~14 reporting flow against a fabricated actively-exploited CVE, tracing it through the chain (CVE $\to$ SBOM component $\to$ sub-asset $\to$ control $\to$ CRA objective) to notification drafting and ENISA single-reporting-platform submission, well inside the 24-hour early-warning clock. The rehearsal exposed workflow, not technical, gaps (a single-owner ENISA-portal account, a missing customer-communications template, an unwritten legal sign-off path), all closed that week and now in the SEUXDR runbook.

\section{Discussion}\label{sec:discussion}
\subsection{Key Findings}\label{sec:lessons}
Three findings recur. First, \emph{documentation gaps dominate over implementation gaps}: the controls usually exist, but policies are unapproved, evidence is not indexed, and SLAs are unwritten, so the GRC engineer's priority is policy approval and evidence collation, not new controls. Second, \emph{AI adds irreducible control overhead}: once an LLM sits in the active-response loop, the ten SCF~AAT controls become non-optional, and a kill-switch test, a rollback playbook, and an explicit boundary on unattended model actions must all be evidenced, or an auditor will probe the emergent-behaviour, value-drift, and human-in-the-loop gaps that generic IAM and change-management controls miss. Third, \emph{reporting readiness is a workflow problem, not a technical one}: the Article~14 24-hour and 72-hour clocks are unforgiving, and our dry-run failures were missing portal credentials, an unclear escalation owner, and no notification template, so a rehearsal before the clock starts is mandatory.

Underlying all three is \emph{scope discipline}: the Step~1 product boundary drives every downstream artefact, so mid-pilot scope drift was our most expensive failure mode, and locking the boundary behind engineering, product, and legal sign-off is cheap insurance. The internal declaration of conformity is likewise best drafted once scope is signed off, since legal review is sequential and rarely shortens.

\subsection{Advantages}\label{sec:advantages}
The qualitative gains over the team's spreadsheet-and-free-form-draft baseline are threefold. (i) \emph{A single source of truth}: the compliance chain links risks, controls, policies, evidence, and CRA objectives in one navigable graph the baseline lacked. (ii) \emph{One-click report export}: the audit pack and EU declaration-of-conformity draft are generated from the chain, not reassembled by hand. (iii) \emph{The chain doubles as an incident-triage tool}: when a CVE drops, the on-call engineer pivots from CVE through SBOM component, sub-asset, controls, and CRA objective to the Art.~14 decision in a few clicks. These are the conformity capabilities (\S\ref{sec:related}) a detection tool alone does not provide.

\subsection{Limitations}\label{sec:limitations}
The study captures what a small manufacturer team produces internally before any external review, a deliberately manufacturer-side scope; formal conformity with a notified body, external-auditor sign-off, customer-side validation, and a real (not tabletop) Article~14 cycle are out of scope, and effort and timing claims are self-reported. External validity is bounded by the single-case design (one product, one manufacturer, one country): CYBERFORT's mapping library reflects its seed data, so other GRC tools yield different mappings, and the AI control family is sensitive to model and playbook scope, so substitutions shift the control set and residual-risk profile. We thus expect the recipe (\S\ref{sec:methodology}) to transfer, but not the absolute numbers (\S\ref{sec:snapshot}). Matching product controls to CRA objectives demonstrates preparedness, not a guarantee of conformity; only a notified-body route can establish the latter.

\subsection{Future Work}\label{sec:future}
Follow-up work extends the study along four axes: multi-team, multi-product replication with instrumented time-on-task; an external-auditor walkthrough of the audit pack against an entry-level conformity-readiness bar; a structured customer survey on the operator surface; and a live Article~14 reporting cycle, with retrospective timings, on the next vulnerability disclosed in the SEUXDR supply chain. We also plan to formalise the Coordinated Vulnerability Disclosure channel Annex~I Part~II obliges---disclosure channel, advisory pipeline, researcher-acknowledgement policy, and customer-communications boilerplate---reusing the ISO/IEC~29147~\cite{iso29147} and ISO/IEC~30111~\cite{iso30111} templates, and to regenerate the SBOM on every release rather than every audit, tracked against the CRA support period.

\section{Conclusion}\label{sec:conclusion}
We documented a six-step, reproducible recipe for CRA conformity preparation and applied it end-to-end to SEUXDR on the open-source CYBERFORT platform. The findings (\S\ref{sec:lessons}) condense into one practitioner message: for engineering-led SMEs the binding constraint is the documentation, traceability, and reporting workflow rather than missing controls, and an AI-augmented product adds an irreducible control overhead once an LLM sits in the active-response loop. The traceability matrix behind these claims is released alongside the open-source code, and the platform itself is open to any SME after free registration, so that the recipe can be reproduced end-to-end without bespoke setup.

\section*{Acknowledgment}
This work has received funding from the Digital Europe Programme under Grant Agreement No.~101190281 (CYBERFORT).

\end{document}